\documentclass[conference]{IEEEtran}
\IEEEoverridecommandlockouts
\usepackage{cite}
\usepackage{amsmath,amssymb,amsfonts}
\usepackage{algorithmic}
\usepackage{graphicx}
\usepackage{tabularx}
\usepackage{attachfile}
\usepackage{textcomp}
\usepackage{float}
\usepackage{xcolor}
\usepackage{tcolorbox}

\tcbset{
    highlightbox/.style={
        colback=yellow!20, 
        colframe=white!75!black, 
        coltitle=black, 
        fonttitle=\bfseries, 
        boxrule=0.5mm, 
        arc=4mm, 
        outer arc=0mm,
        boxsep=5pt, 
        left=5pt, 
        right=5pt, 
        top=5pt, 
        bottom=5pt, 
    }
}

\newtcolorbox{boxA}{
    fontupper = \bf,
    boxrule = 1.5pt,
    colframe = black
}

\def\BibTeX{{\rm B\kern-.05em{\sc i\kern-.025em b}\kern-.08em
    T\kern-.1667em\lower.7ex\hbox{E}\kern-.125emX}}
\begin{document}

\title{Comparing Human and LLM Generated Code: The Jury is Still Out!\\
}


\author{
\IEEEauthorblockN{Sherlock A. Licorish}
\IEEEauthorblockA{\textit{School of Computing}\\
\textit{University Of Otago}\\
New Zealand\\
sherlock.licorish@otago.ac.nz}
\and
\IEEEauthorblockN{Ansh Bajpai}
\IEEEauthorblockA{\textit{Faculty of Information Technology}\\
\textit{Monash University}\\
Australia\\
anshbajpai8@gmail.com}
\and
\IEEEauthorblockN{Chetan Arora}
\IEEEauthorblockA{\textit{Department of Software Systems \& Cybersecurity}\\
\textit{Monash University}\\
Australia\\
chetan.arora@monash.edu}
\and
\IEEEauthorblockN{Fanyu Wang}
\IEEEauthorblockA{\textit{Department of Software Systems \& Cybersecurity}\\
\textit{Monash University}\\
New Zealand\\
fanyu.wang@monash.edu}
\and
\IEEEauthorblockN{Kla Tantithamthavorn}
\IEEEauthorblockA{\textit{Department of Software Systems \& Cybersecurity}\\
\textit{Monash University}\\
Australia\\
chakkrit@monash.edu}}

\maketitle

\begin{abstract}
Much is promised in relation to AI-supported software development. However, there has been limited evaluation effort in the research domain aimed at validating the true utility of such techniques, especially when compared to human coding outputs. We bridge this gap, where a benchmark dataset comprising 72 distinct software engineering tasks is used to compare the effectiveness of large language models (LLMs) and human programmers in producing Python software code. GPT-4 is used as a representative LLM, where for the code generated by humans and this LLM, we evaluate code quality and adherence to Python coding standards, code security and vulnerabilities, code complexity and functional correctness. We use various static analysis benchmarks, including Pylint, Radon, Bandit and test cases. Among the notable outcomes, results show that human-generated code recorded higher ratings for adhering to coding standards than GPT-4. We observe security flaws in code generated by both humans and GPT-4, however, code generated by humans shows a greater variety of problems, but GPT-4 code included more severe outliers. Our results show that although GPT-4 is capable of producing coding solutions, it frequently produces more complex code that may need more reworking to ensure maintainability. On the contrary however, our outcomes show that a higher number of test cases passed for code generated by GPT-4 across a range of tasks than code that was generated by humans. That said, GPT-4 frequently struggles with complex problem-solving that involve in-depth domain knowledge. This study highlights the potential utility of LLMs for supporting software development, however, tasks requiring comprehensive, innovative or unconventional solutions, and careful debugging and error correction seem to be better developed by human programmers. We plot an agenda for the software engineering community.\\

\end{abstract}


\section{Introduction}
\label{sec:introduction}
The incorporation of artificial intelligence (AI) into software development, notably through large language models (LLMs) such as GPT-4, Gemini, LLaMA, Claude and Mistral, is a disruptive trend that is set to revolutionise software development~\cite{b21}. These models promise to speed up software development time, minimise human error, and improve coding efficiency~\cite{b22}. However, with increasing use of these technologies, it becomes critical to systematically assess their efficacy and safety, particularly in comparison with human programmers~\cite{b1}. The idea is that if LLMs are able to match humans, they may be used to at least support them during the software development process. The fact is, even though previous work have made attempts to evaluate the fine-tuning or pretraining performance of LLMs~\cite{b26,b27}, their results do not conclusively support incorporating AI into software development. The limitations could be formulated as following two viewpoints: (i) from a practical perspective, prompt generation needs further consideration while research largely covers fine-tuning or pretraining results; (ii) there is need to evaluate the alignment between human- and LLMs-generated codes to verify how LLMs will support humans during software development.

Thus, in order to provide comprehensive guidelines for supporting software development through LLMs code generation, especially for the participants involved in Python software development, we conduct a systematic quality evaluation of code created by humans versus that generated by LLMs. Unlike previous work evaluating the alignment between human- and LLM-generated code by conducting human interviews or exploring a single quality dimension, our study explores multiple quality dimensions. Specifically, inspired by previous assessments of code health and wholesomeness~\cite{b23}, we design our evaluation to cover: the mitigation of security risks, coding standards compliance, code complexity, and functional correctness, essential code quality dimensions~\cite{b2,b3,b4}. Moreover, to provide detailed results, we examine 72 different Python code-generated files, each representing a different software engineering task.

The evaluation dimensions are quantitatively investigated in this study, with subsequent follow up deeper contextual examination. \textit{Security}, which is one of the primary quality factors, remains a significant concern during software development~\cite{b24}. That said, LLMs, by definition, generate code based on patterns learned from large datasets, which often contain coding flaws~\cite{b6,b7}. This paper carefully examines the security implications of code generated by LLMs, identifying typical security flaws and comparing these flaws to those found in human-written code, including cross-site scripting (XSS), and other vulnerabilities. In addition to security flaws, this study examines \textit{coding standards} using linting (static analysis) tools~\cite{b25}. Linting tools facilitate the examination of code for potential flaws and bad practices, helping to uncover variations in coding standards adherence and flagging sub-optimal practices that may lead to future errors. Comparing the outcomes from human- and LLM-generated code, we assess LLMs' readiness and reliability for software development~\cite{b8,b9}. As for the \textit{functional correctness} assessment, human-written and LLM-generated coding outcomes are tested using the same unit test cases. This testing allows a direct comparison of how effectively each codebase completes the intended tasks, providing deeper insight into their functional performance and reliability~\cite{b23}. \textit{Code complexity} is the last factor considered. Using the cyclomatic complexity metric, we count the number of linearly independent pathways across source code~\cite{b23}. This metric helps determine the testability and maintainability of software. A comparison of cyclomatic complexity between human- and LLMs-generated code provides an understanding of how the usage of LLMs may affect the overall complexity of software and future maintenance efforts.

The consideration of the aforementioned comprehensive factors enables insightful analysis of the human-LLMs alignment of code generation. Generally, the contributions provided by this study are twofold: (i) under real workflow scenarios in software development, we conduct a comprehensive evaluation of the results of human- and LLMs-generated codes, which shows the gaps and limitations in current code generation methods; (ii) we provide recommendations for how the software engineering community may incorporate AI into software development, which serves as guidelines for improving productivity and software outcomes. 

This paper is organised as follows. Section II provides the study background and reviews related work. Section III introduces our study settings, including our data and methods, measures and tools, and data analysis. Section IV reports our results, where we analyse the outcomes. Section V summarises the implications of our findings, and Section VI considers threats to the study. We provide concluding remarks in Section VII, before outlining future work in Section VIII.

\section{Background}
\label{sec:literature}

Significant changes have been sparked by the integration of Large Language Models (LLMs) into software development, especially with regard to the automation of coding procedures~\cite{b21,b22,b40}. This literature review examines a number of LLM capabilities, including how well they generate code, how they perform compared to human-generated code, and the security and ethical implications of LLMs' involvement in code generation.

The capacity of LLMs to produce text that resembles that of a human has significantly increased recently, opening up applications for the technology beyond basic text generation to encompass complex jobs like developing code. Code development productivity can be increased and error rates can be decreased by automating typical programming operations, as demonstrated by the development of models such as Gemini, Claude, and GPT-4~\cite{b1,b2}. These models make use of huge amounts of data to carry out a variety of functions and have been shown to be capable of understanding and generating code snippets, offering contextual recommendations, and automating debugging procedures~\cite{b3}.

That said, the quality of code generated by LLMs (like humans) is paramount, with studies emphasizing the need for robust metrics to evaluate aspects such as maintainability, readability, and performance~\cite{b7,b17}. Tools like Radon and Pylint have been instrumental in quantifying these metrics, providing valuable insights into the structural quality of code. For example, Radon measures cyclomatic complexity, offering a way to understand code complexity, while Pylint evaluates adherence to coding standards and identifies potential errors~\cite{b4, b5}. These tools play a crucial role in ensuring that LLM-generated code meets high standards of quality and maintainability. In addition, specialized tools and approaches have been developed to enhance code evaluation. Ren et al.\cite{b35} introduced CodeBLEU, an automated evaluation tool for code synthesis, which adapts evaluation metrics from the NLP domain to assess code generation. Similarly, Nguyen et al.\cite{b36} evaluated the readability and understandability of code generated by Copilot on test problems from LeetCode. Recently, there has been a growing interest in assessing code quality from practical or human-centric perspectives. For example, EvalPlus is a comprehensive empirical method for evaluating code quality~\cite{b33}, considering multiple factors such as functional correctness, test adequacy, code coverage, mutant kill rates, empirical LLM sample failures, error detection, and ranking adjustments. Wang et al.~\cite{b34} conducted an experimental evaluation of two software engineering tasks: code puzzles and typical software development challenges. By involving human participants however, our study incorporated a human-centric dimension to evaluate the helpfulness of LLM-generated code in real-world software engineering workflows, thereby emphasizing practical usability alongside technical quality.

Security, in particular, is still a major concern for human- and LLM-generated code. It is vital to concentrate on creating methods for vulnerability evaluation and mitigation because LLMs have the power to spread unsafe coding practices if they are trained on data with coding vulnerabilities. Because LLMs are increasingly being used in software development~\cite{b21,b22,b26,b27}, static analysis techniques like Bandit are essential for finding common security issues~\cite{b6}. Research has demonstrated that LLMs may unintentionally introduce security flaws such as hardcoded secrets and inappropriate handling of private information, underscoring the necessity of implementing strong security mechanisms~\cite{b7}.

Notwithstanding potential biases~\cite{b8} and practical challenges~\cite{b9}, including stereotypes around gender~\cite{b29} and culture~\cite{b28} that LLMs may enforce, investigations exploring and comparing human- and LLM-generated code would provide insights into how well LLMs perform in comparison to human programmers. However, such studies remain scarce, with most research focusing on how humans utilize LLMs during coding~\cite{b30,b31} or on the accuracy of LLMs in solving specific programming tasks~\cite{b26,b27}. There are a few exceptions, for example, Prather et al.~\cite{b37} performed a contrastive study between human- and LLM-generated code using human interviews. In this study, participants were presented with pairs of code samples and asked to answer questions regarding understandability, preferences, advantages, and concerns. However, this approach relies heavily on subjective human evaluations and lacks rigorous comparison results. 

In contrast, Awal et al.~\cite{b38} designed an experimental evaluation to compare human-written code with code generated by BERT and CodeGPT. This study primarily examined the security aspect by introducing an adversarial attack strategy to evaluate code robustness. Nonetheless, the study has notable limitations: (i) a narrow focus on evaluation factors, as security is not the only aspect requiring consideration, and (ii) the adoption of BERT and CodeGPT, which are not state-of-the-art (SOTA) LLMs for code generation. While LLMs are good at writing code for clearly described issues, they frequently struggle with more difficult tasks that call for detailed understanding or original problem-solving techniques. For instance, Hochmair et al.'s~\cite{b12} study shows that while LLMs are effective at spatial reasoning tasks, they are less successful in more complex scenarios needing in-depth contextual knowledge. That said, questions remain such as: How do LLMs compare to human programmers in terms of code security, quality and efficiency within software development tasks?, In which areas does LLM-generated code exhibit weakness or underperformance when compared to human-written code, particularly in terms of task execution, complexity management, and error handling?

This work addresses this opportunity, where detailed evaluations are conducted to compare human- and LLM-generated code.

\section{Study Settings}
\label{sec:methods}
\subsection{Data and Methods}

In this study we use GPT-4 (OpenAI's GPT-4 that was available in the first two weeks of April 2024) as a representative LLM for comparison between LLM- and human-generated Python code across various performance metrics. This LLM was held to be the best performing model at the time of our experiments. We use the quantitative comparative analysis method  for systematically comparing various measurable characteristics of code produced by LLMs and human programmers~\cite{b18,b19}. This design was chosen because it enables objective evaluations. Radon~\cite{b5}, Pylint, and Bandit\cite{b6} were used for code evaluation, as they have been widely acknowledged and verified for their efficacy and reliability~\cite{b15,b17,b4}.

The study uses a benchmark dataset of 72 Python coding tasks, covering a variety of software engineering problems sourced from a previous study~\cite{b7}. A fourth year computing student with software development experience (human  programmer) developed the code for the 72 coding tasks, while code samples were created using GPT-4 (i.e., the version available in April 2024) in order to produce equivalent LLM output. The full list of tasks and associated solutions are available on request. We used zero-shot prompting to guarantee that every prompt was consistent and fair for comparison~\cite{b10}. While it is possible to repeatedly iterate prompting (i.e., using few-shots and chain-of-thought prompting techniques) in an effort to engineer improvements in solutions generated by LLMs~\cite{b39}, we believe such experimental configurations would not fairly support comparisons as planned in the work (i.e., human-generated solutions were not evolved after repeated evaluations). Test files were created for the human- and LLM-generated code, where testing circumstances were consistent for the 72 tasks and evaluations were reliable and fair.

\textbf{Prompting:} In developing each prompt, contextual data was given in an effort to improve the LLM's output/responses. This contextual setup contained brief overviews of the task's requirements, expected outcomes, and any particular coding guidelines that had to be followed. An example is shown below. 

\begin{itemize}
    \item Original Data Source
    \begin{itemize}
        \item Taken from the original dataset~\cite{b7}, a typical example task involves "creating a Python function to approximate the sine function using a particular series expansion". Our new dataset comprises the task, human- and LLM-generated solutions.\
    \end{itemize}
    \item Prompting for LLM-Generated Code
    \begin{itemize}
        \item Prompt: "Write a Python function that approximates the sine function using the Taylor series expansion. Include error handling for invalid inputs and document the code with comments explaining the logic."\
        \item GPT-4 was charged with producing code based on this prompt using zero-shot prompting. The prompt was designed to guarantee that the code generated would follow best standards for coding, including error handling and documentation, in addition to completing the task in question. One LLM-generated solution was saved for evaluation. 
    \end{itemize}
    \item Test Cases
    \begin{itemize}
        \item For the scenario (solution generated) above, multiple test cases were designed to pass a set of angles to the function and check if the returned values are within a reasonable margin of the true sine values, as follows.\
        \begin{align*}
            0, \frac{\pi}{6}, \frac{\pi}{4}, \frac{\pi}{2}
        \end{align*}
        \item To provide consistency in testing, the same test cases were executed for both human- and LLM-generated code. Tests were created for all of the 72 tasks.\
    \end{itemize}

\end{itemize}

\subsection{Measures and Tools}
An automated Python script was employed to gather data for this study. It was created to execute and capture data for 72 different Python code files, which represented a variety of software tasks. Industry-standard technologies such as Pylint for code quality evaluation, Radon for code complexity assessment, and Bandit for security analysis were used in the evaluation process. Additionally, each code snippet's functional correctness was tested using the Pytest package. The use of such an approach for studying code quality is established~\cite{b25}. We provide additional details on the measures below.
\begin{itemize}
    \item \textit{Pylint}: Used to assess/measure code quality and code adherence or compliance to Python coding standards. Code is given a rating by Pylint on a range of -10 to 10. Scores nearing 10 signify exceptional compliance with Python coding guidelines~\cite{b20} and general code well-being, whilst lower scores can suggest possible coding violations or usage of inadequate practices.\

    \item \textit{Radon}: Employed for measuring code complexity. With respect to cyclomatic complexity, Radon assigns grades ranging from `A' (most maintainable) to `F' (least maintainable) and provides an average complexity score of the code in a float format, which is utilised in our research. Lower scores correspond to less maintainable and more complex code.\

    \item \textit{Bandit}: Utilized for static security analysis to identify common vulnerabilities within code. Bandit reports problems on three tiers of severity: LOW, MEDIUM, and HIGH. Additionally, we have aggregated the total issues generated by a coding solution (file) to represent the codebase's overall security state, where a lower score denotes fewer vulnerabilities.\

    \item \textit{Pytest}: Applied to execute the written test cases, evaluating the functionality and correctness of both human- and LLM-generated code. Pytest results are typically binary, represented as PASS or FAIL for each test case. The percentage of passed tests was also used to quantify the functional correctness or executability of the code.
\end{itemize}

\subsection{Data Analysis}
To enable a thorough and in-depth analysis of the performance differences between human- and LLM-generated code, two main kinds of result files were carefully generated. Under the same evaluation criteria, these files are essential for understanding the strengths and weaknesses of both sources. A brief description of each is provided below:
\begin{itemize}
    \item Statistics-based Metrics: Metrics collected above offer a measurable assessment of the performance of human- and LLM-generated code in terms of functional correctness, code complexity, security vulnerabilities, and compliance to coding standards. These metrics were analysed using various statistical procedures (e.g., box plots and formal statistical testing).\

    \item Detailed Issues: All the issues found during the evaluation of each code file are manually explored by at least one author to complement the statistical analysis and provide detailed context around the faults in human- and LLM-generated code. This helps us investigate in detail the kinds of mistakes that are common in human- and LLM-generated code, highlighting particular instances in which LLMs perform better/worse than human programmers~\cite{b32}. Such deeper analysis also supports us in making inferences and teasing out implications for future use of LLMs for code generation.
\end{itemize}

\section{Results and Analysis}
\label{sec:results}

\subsection{Code Compliance to Python Coding Standards}
As noted above, Pylint checks Python code for mistakes and compliance with coding standards. Pylint was utilised in our research to evaluate the documentation, coding conventions, and error detection quality of Python code generated by humans and LLMs. We discuss outcomes for these dimensions in turn.\

\subsubsection{Missing Docstrings and Documentation Standards}
\begin{itemize}
    \item Outcome: Outcomes revealed a notable lack of module and function docstrings for code generated by both humans and LLMs. For instance, for complicated jobs like the tic-tac-toe GUI implementation and pandemic simulation, 85\% of the GPT-4 outputs lacked complete docstrings. Similarly, this problem was present in 78\% of human-written code.\

    \item Analysis: The lack of documentation in both human- vs LLM-generated code suggests there is need for improvement in both domains, especially for code likely requiring high maintainability and where new programmers may need to interface with the code.\
    
\end{itemize}

\subsubsection{Coding Standard Violations}
\begin{itemize}
    \item Outcome: Variable and constant naming violations were observed on a large number of occasions (72 in GPT-4 code and 65 in human code), particularly in the application of scientific algorithms and game logic.\

    \item Analysis: Violating naming conventions can cause confusion for programmers and slow down development. Outcomes here indicate that although LLMs make more mistakes in applying these standards than humans do, both humans and LLMs struggle with adhering to naming conventions.\
    
\end{itemize}

\subsubsection{Redefinition of Variables (Outer Scope)}
\begin{itemize}
    \item Outcome: It was observed that 45\% of GPT-4 generated code and 30\% of human code had incorrect redefinitions from outer scope reported by Pylint, especially in scripts with nested functions or complicated loops.\

    \item Analysis: Redefinitions of this kind may introduce risks related to variable scope. Future LLM model training need to address this issue, in enhancing contextual awareness rate in LLM-generated code. That said, human generated code is not perfect, and also needs vigilance.    
\end{itemize}

\subsubsection{Import and Structure Issues}
\begin{itemize}
    \item Outcome: Results show that 55\% of GPT-4-generated files had errors related to import ordering and missing dependencies, particularly in files that needed third-party libraries, such as bpy, for Blender tasks. In contrast, such problems were present in only 40\% of the human generated code.\

    \item Analysis: Outcomes here illustrate the difficulty with managing dependencies and library-specific situations in code generated by LLMs, which could be problematic in deployment circumstances where reliable and error-free execution is necessary. At the same time, human generated code is not perfect, albeit better than code created by LLMs. 
\end{itemize}

\begin{figure}[h]
  \centering
  \includegraphics[width=0.5\textwidth]{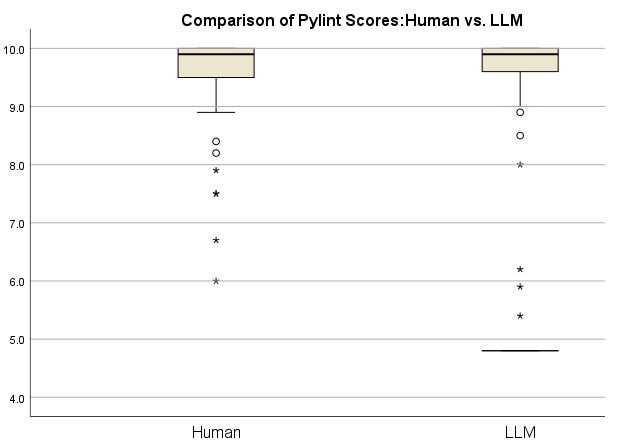}
  \caption{Comparison of Human vs LLM Pylint Scores in box plots}
  \label{fig:Pylint-Box}
\end{figure}

We provide box plots of PyLint scores for human- and LLM-generated code in Fig. \ref{fig:Pylint-Box}, which show that human-generated code on average receives slightly higher ratings for adhering to coding standards. LLM-generated code, on the other hand, shows more variation; where we observed ratings that range from 10 to as low as 4.8. This variation points to inconsistent documentation handling practices or coding standard adherence by LLMs. Also, with a wider interquartile range for LLM-generated code compared to human-generated code, the box plots highlight the frequent irregularity in addition to overall slightly lower code quality for LLM-generated code. That said, statistical testing involving t-tests to compare the mean difference in PyLint scores of human- and LLM-generated code did not reveal statistically significant differences ($p=0.64$).\

\begin{boxA}
While LLMs are capable of producing syntactically accurate code, when compared to human programmers, GPT-4 falls short of recommended practices at times and exhibits variation in code quality and context awareness when assessed with Pylint. The practical effectiveness of LLM-generated code may be increased by improving training procedures, such as combining a variety of coding patterns and enforcing stronger coding standards in training data.
\end{boxA}

\subsection{Code Security and Vulnerabilities}
Python code is parsed by Bandit to identify security flaws such as insecure library usage, unsafe subprocess handling, hard coding of sensitive data, and other behaviours that increase code vulnerabilities and lead to security breaches. As noted above, three severity levels are distinguished by the tool: low, medium, and high. To enable targeted remediation, each issue is further coded (e.g., B601, B602) according to the type of vulnerability present.

\begin{figure}[h]
  \centering
  \includegraphics[width=0.5\textwidth]{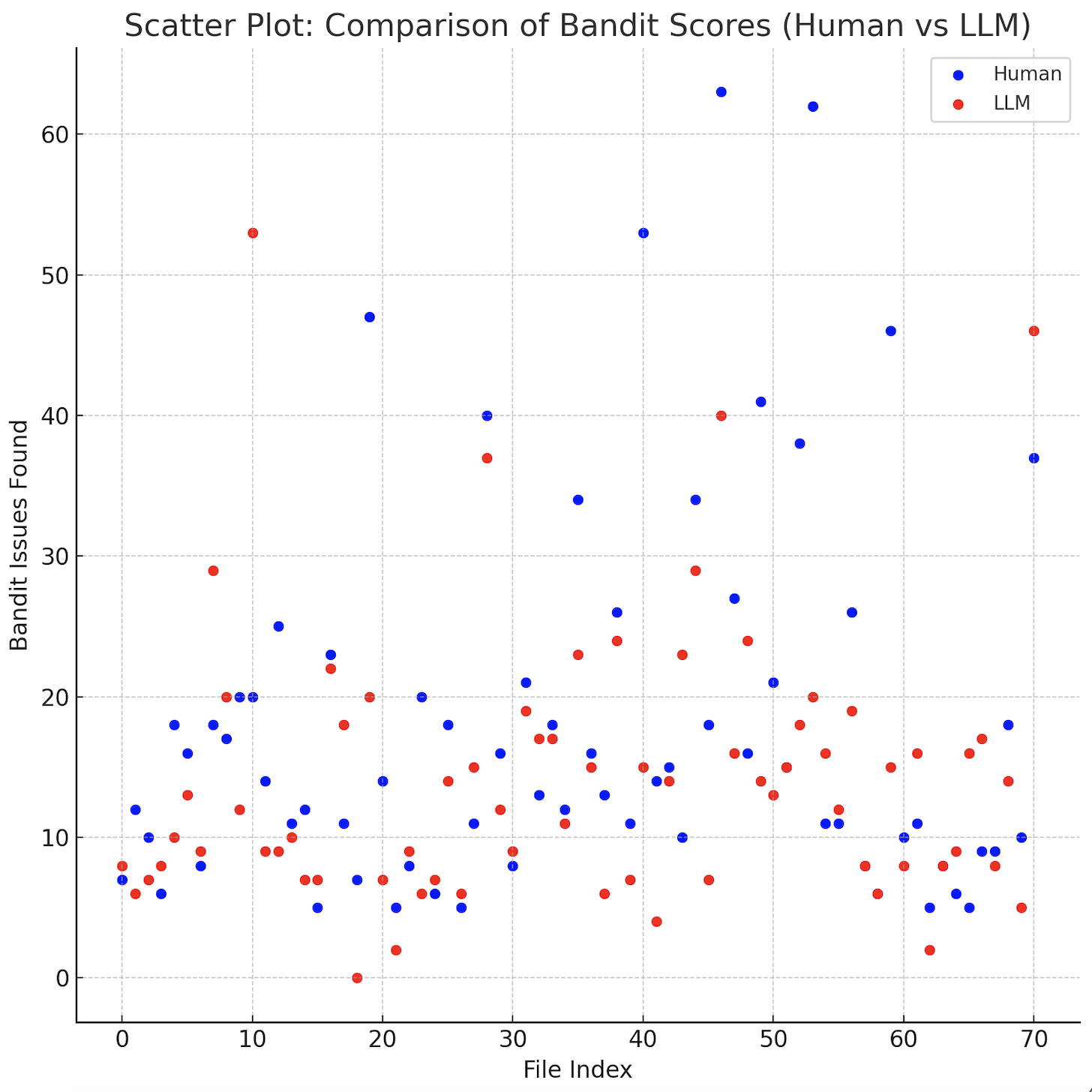}
  \caption{Comparison of Human vs LLM Bandit Scores in a scatter plot}
  \label{fig:Bandit-Scatter}
\end{figure}

\subsubsection{Findings and Analysis Across Files}
\begin{itemize}
    \item \textit{Subprocess Vulnerabilities (High Severity)}: Subprocesses were commonly handled incorrectly by both human and GPT-4 code, which is a serious flaw and can result in arbitrary code execution. For example, code instructions meant for automation tasks where commands were created dynamically from user inputs were frequently found to have the Bandit flag B603 enabled. In another example of a serious flaw, a script generated dynamically created shell commands without any sanitation based on external inputs, potentially opening the door to command injection vulnerabilities. In fact, in GPT-4-generated code, 40 instances of dangerous subprocesses usage were observed, mostly in scripts meant for file handling and system command executions.

    \item \textit{Hard-coded Sensitive Data (Medium to High Severity}: Several scripts were found to have hard-coded passwords and API keys for both human and GPT-4 code, which increases the possibility of credential leakage. This was especially common in scripts managing API connections, where the API keys were code-integrated. If unauthorised individuals are able to access data or if the code is made publicly available, such actions may result in serious security breaches. In fact, there were 25 instances of hard-coded sensitive data (API keys, passwords) across both human-written and LLM-generated scripts.
   
    \item \textit{Insecure Library Usage (Low to Medium Severity)}: Another prevalent issue, B404, includes the usage of known insecure modules like Pycrypto, which are found in various data encryption routines. This issue was also related to the use of redundant or vulnerable libraries. A particular example includes the usage of an old version of the Requests library in LLM-generated code, which is known to have several vulnerabilities that may lead to SSL spoofing attacks. Outcomes revealed 18 instances of such issues in LLM-generated scripts, mainly involving outdated or redundant cryptography libraries.
\end{itemize}

\subsubsection{Further Statistical Comparisons}
\begin{itemize}

        \item High Severity: Compared to 45\% of issues in human-generated code, 60\% of issues in LLM-generated code were of high severity.\

        \item Medium Severity: We observed that 30\% of such issues were found in LLM-generated code, compared to 35\% in human-generated code.\

        \item Low Severity: Findings show that 10\% of low severity issues were in LLM-generated code, compared to 20\% in human-generated code.

\end{itemize}

\begin{figure}[h]
  \centering
  \includegraphics[width=0.5\textwidth]{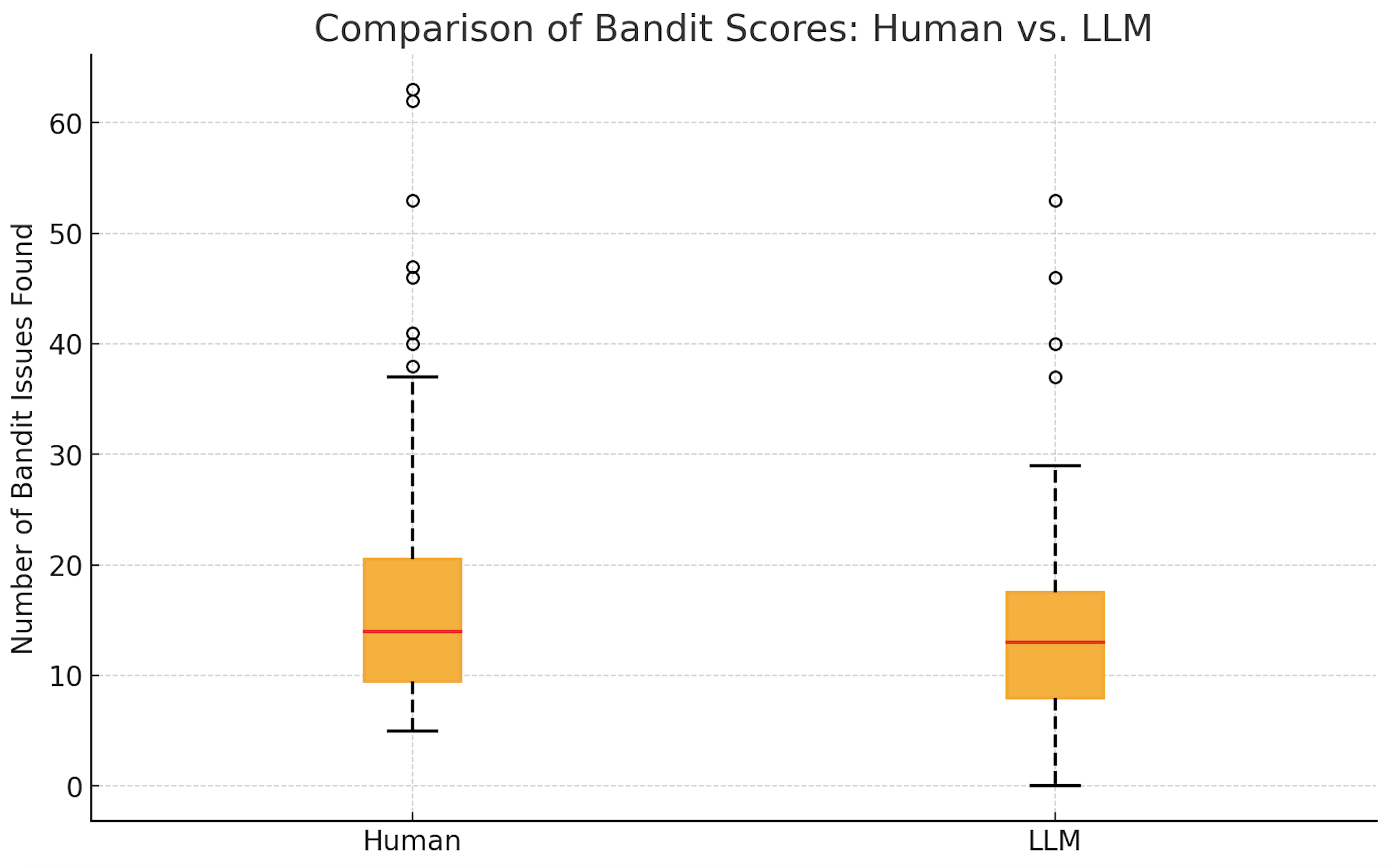}
  \caption{Comparison of Human vs LLM Bandit Scores in box plots}
  \label{fig:Bandit-Box}
\end{figure}

Analysis: Security flaws are common in both human- and LLM-generated code, as demonstrated by the security analysis of Python code above. A wide range of problems are visible in the scatter plot in Fig. \ref{fig:Bandit-Scatter}, with both human and LLM-generated code displaying security breaches. The median vulnerability in the box plot in Fig. \ref{fig:Bandit-Box} for the LLM and human codes are similar, with human-generated code showing more outliers. T-test results returned a p-value of 0.05, which is at the edge of statistically significant differences (typically $p<0.05$). 

\begin{boxA}
Outcomes for security flaws being common in both human- and LLM-generated code, with such flaws in LLM-generated code being more severe, highlight the urgent need for improved security features across the coding divide. Accordingly, a case is made in this study for prioritising security in software development and training LLMs with more secured code.
\end{boxA}

\subsection{Code Complexity}
We used Radon in this study, a Python tool that measures cyclomatic complexity. Cyclomatic complexity is an objective measure of a program's complexity and maintainability that counts the number of linearly independent pathways through the source code\cite{b25}. As part of our code quality measures, we examine the cyclomatic complexity of the 144 (72 x 2) different Python coding files generated by humans and LLMs.

\subsubsection{Findings and Analysis Across Files}
\begin{itemize}
    \item \textit{General Cyclomatic Complexity}: The average code complexity of the code generated by GPT-4 is 5.0, 61\% higher than that of human code (average code complexity = 3.1). This may be due to the model's tendency to handle more edge scenarios or include extra checks and balances than would be included by programmers. In fact, a generalised approach to code generation seems to be evident for LLM-generated code, with a code complexity range between 4.5 and 5.4 (standard deviation = 0.2). This may lead to overfitting solutions that anticipate more scenarios than are actually necessary. Also, LLMs may be trained on a variety of coding styles and solutions. On the contrary, code produced by humans seems to be of limited variety, with the range for code complexity being 2.69 to 3.91 (standard deviation = 0.4). This latter outcome indicates that humans typically employ similar levels of abstraction and control flow structures when solving different coding tasks.
    
    \item \textit{Task-Specific Cyclomatic Complexity}: The cyclomatic complexity of coding tasks performed by LLMs, such as Sudoku solving, appeared to be higher. This could be because LLMs provide more comprehensive responses that cover a wider range of scenarios and incorporate extra validation and error-checking procedures. For scripting tasks, LLMs created more complex code for tasks where simple solutions are typical, such as data parsing and JSON POST requests. This implies a built-in bias in the model towards thorough, occasionally over-engineered solutions.
\end{itemize}

\subsubsection{Further Analysis of Impact}
\begin{itemize}
    \item The inclination for LLM-generated code to be more complex could make code review procedures and long-term software maintenance difficult. To assure reliability and maintainability, more thorough testing and review cycles might be required. As code complexity increases, there is a greater chance that errors or bugs will be missed during testing. In circumstances where reliability is crucial, this could have serious implications for software quality.\ 
\end{itemize}

\begin{figure}[h]
  \centering
  \includegraphics[width=0.5\textwidth]{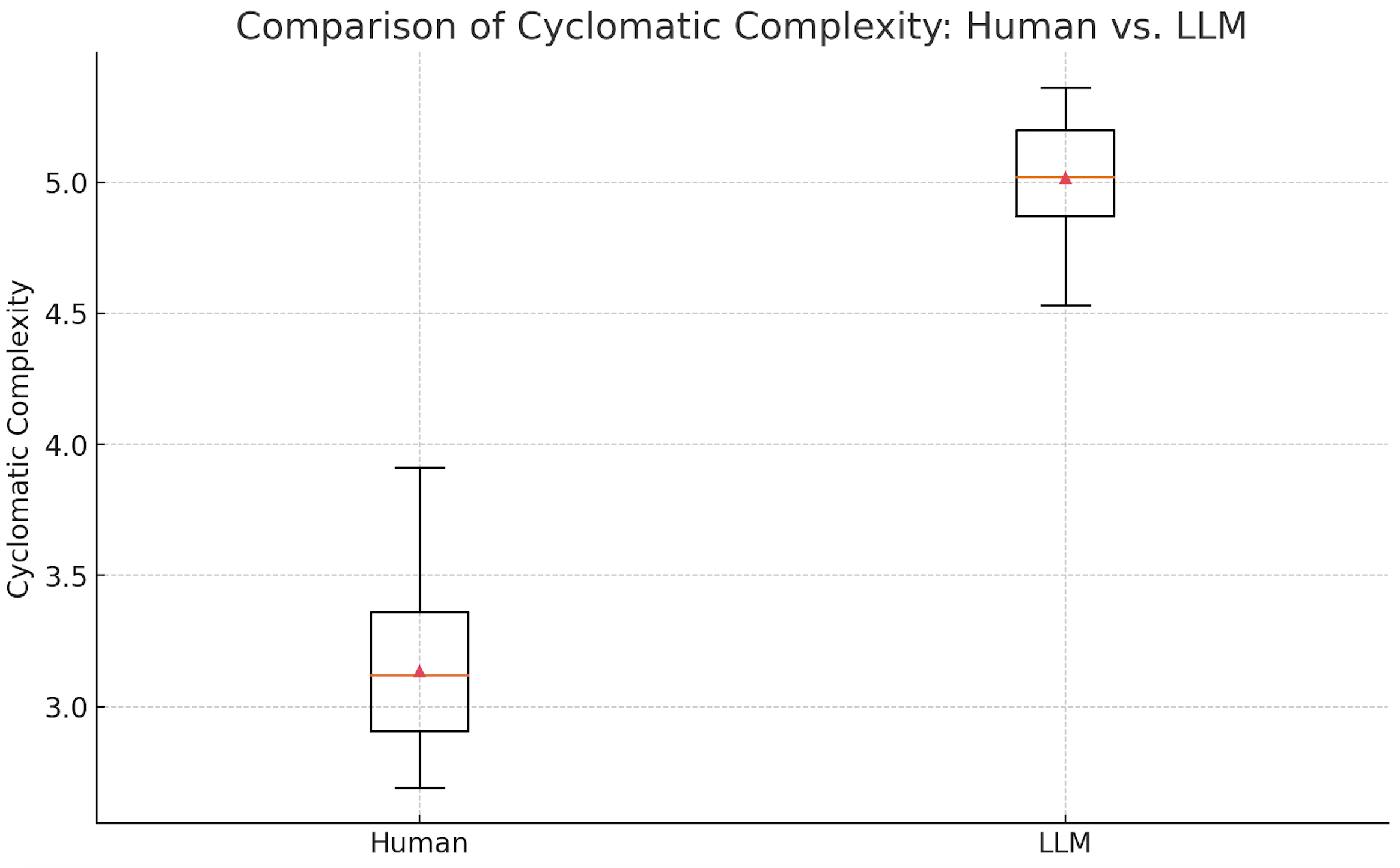}
  \caption{Comparison of Human vs LLM Radon Scores in a box plots}
  \label{fig:Radon-Box}
\end{figure}

Fig. \ref{fig:Radon-Box} shows the box plots for human- and LLM-generated code, highlighting the differences in cyclomatic complexity. Here, it is shown that, on average, LLMs generate code that is more complex, as seen by a higher median and a wider interquartile range. Formal statistical testing using the T-test also confirms statistical differences for human- and LLM-generated code ($p<0.05$).
Our results show that although LLMs are capable of producing coding solutions, they frequently produce more complex code that may need more reworking to ensure maintainability.

\begin{boxA}
Although LLMs may increase productivity by producing code drafts quickly, our outcomes emphasise how crucial it is to assess LLM-generated code for complexity problems prior to integrating it into production systems in order to maintain its scalability and performance, especially during software maintenance.\
\end{boxA}

\subsection{Code Functionality and Correctness}
We explore how well human- and LLM-generated code perform in a range of functional software testing conditions. We compare pass/fail rates and details of the types of errors encountered to assess the correctness of the code generated by both sources under carefully controlled testing circumstances. These outcomes are presented and analysed below.

\subsubsection{Pass and Failure Analysis}
\begin{itemize}
    \item \textit{Pass Rate Analysis}: The first step in the research was to compute pass rates, a key indicator of the code's capacity to satisfy functional requirements. The human-generated code pass rate was found to be 54.9\%, and the LLM-generated code pass rate was 87.3\%. This notable difference emphasises how well the LLM performs in test scenarios, indicating that code generated by LLMs may adhere to functional specifications more closely.\

\begin{figure}[H]
  \centering
  \includegraphics[width=0.5\textwidth]{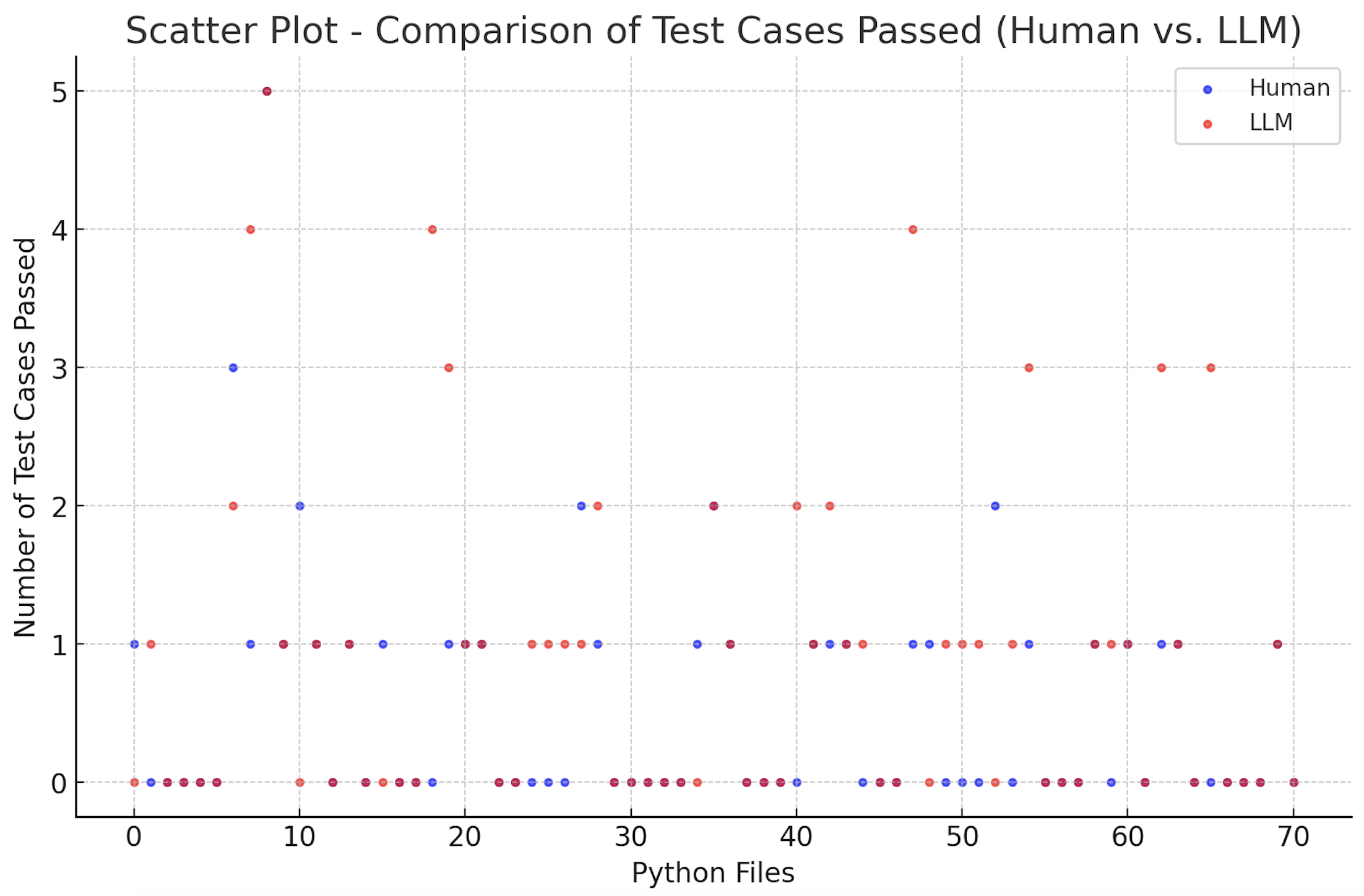}
  \caption{Comparison of Human vs LLM Test Scores in a scatter plot}
  \label{fig:Test-Scatter}
\end{figure}

    \item \textit{Failure Analysis}: In terms of the reasons for failure, human-written code demonstrated a standard deviation of 16.4 among the non-executable scripts, with an average of 21.6 Bandit problems recorded per script. This variation highlights how inconsistently human-written code adheres to best practices. LLM-generated code that failed test cases (i.e., non-executable scripts) had returned a standard deviation of 8.6, with an average of 15.9 Bandit problems.\
    
\end{itemize}

\subsubsection{Detailed Code Analysis}\

While LLM-generated code shows a high pass rate for the 72 test cases, in certain areas human programmers tended to do better than LLMs. These differences are especially noticeable for programming tasks that require sophisticated problem-solving techniques, in-depth subject knowledge, or complex thinking. We reflect on these outcomes below. 

\begin{itemize}
    \item \textit{Complex Logic Concepts}: The task to solve the Travelling Salesman Problem using the Quantum Optimisation Algorithm required knowledge of optimisation methods and quantum computing principles. When comparing the human (pass rate = 75\%) programmers' solutions that made use of domain-specific knowledge and creative problem-solving abilities and LLMs' (pass rate = 35\%) outcomes, it appeared that LLMs found it difficult to integrate these complicated concepts to provide correct, coherent solution. Similarly, tasks involving accurately reflecting on real-world dynamics in simulation activities require both an understanding of the underlying models, such as SIR models in epidemiology, and the flexibility to modify these models. Compared to models created by humans (pass rate = 80\%), LLMs (pass rate = 45\%) produced models that were less accurate or oversimplified. This may be because LLMs do not fully understand the internal details of the model parameters or the implications of particular simulation settings.\

    \item \textit{Context Awareness}: Tasks to interface with an LLM in Python were impacted by context awareness. Interestingly, coding an LLM within a script generated by an LLM frequently reveals the meta-level knowledge that human programmers possess. An LLM's (pass rate = 50\%) training data could not accurately reflect an individual's grasp of AI and machine learning structures, which allows humans (pass rate = 90\%) to critically assess and refine design choices. Similarly, this issue was observed for a task to analyze sentiment from images using GCP. For LLMs (pass rate = 30\%), tasks that call for unique approaches frequently pose challenges. It is generally easier for humans (pass rate = 70\%) to have the creative and contextual adaptation required for certain coding tasks.\

    \item \textit{Debugging and Error Handling}: Fixing a bug in the anagram check code that was provided revealed interesting details. In order to effectively debug an issue, it is necessary to not only recognise its existence but also understand its cause and the best way to address it without creating new mistakes. Iterative debugging procedures require an advanced understanding of the code's function and intent, which is where human (pass rate = 82\%) programmers excel and LLMs (pass rate: 45\%) currently struggle.
\end{itemize}

\begin{figure}[h]
  \centering
  \includegraphics[width=0.5\textwidth]{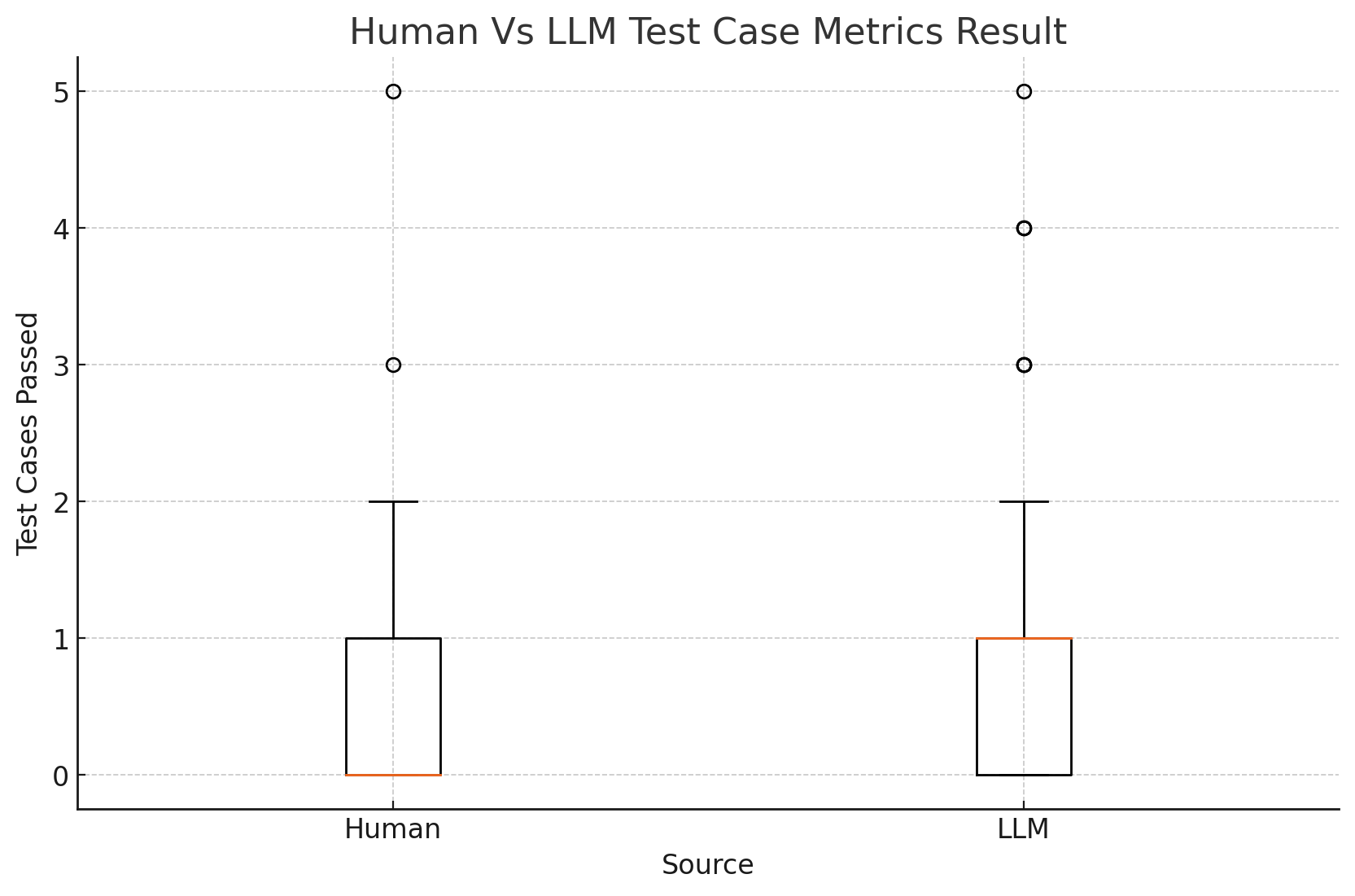}
  \caption{Comparison of Human vs LLM Test Scores in box plots}
  \label{fig:Test-Box}
\end{figure}

We visualise test case outcomes for Python scripts written by humans and generated by LLMs in Fig. \ref{fig:Test-Scatter}, which shows that higher number of test cases passed for LLM-generated code across a range of tasks. This difference is further shown by the box plots in Fig. \ref{fig:Test-Box}, where it is shown that LLMs have a smaller interquartile range, indicating more consistent performance, than does human-generated code. LLMs statistically show an overall pass percentage of 87.3\%, which is much higher than the 54.9\% for human code. Statistical testing also revealed statistically significant differences between human- and LLM-generated code ($p<0.05$). This shows how the LLMs may handle a larger range of cases and include more elaborate error-checking processes. The scatter plot in Fig. \ref{fig:Test-Scatter} indicates that humans do better on tasks requiring more in-depth domain-specific knowledge or innovative problem-solving techniques, even if LLMs are generally more efficient in standard assessments.

\begin{boxA}
In summary, the outcome of this work shows that although LLMs are proficient at a variety of programming tasks, their performance is typically worse in domains demanding in-depth domain-specific knowledge, complex problem-solving techniques, and high context sensitivity. Tasks requiring a comprehensive, natural grasp of complex structures, innovative or unconventional solutions, and careful debugging and error correction seem to be better developed by human programmers.
\end{boxA}

\section{Implications}

\textbf{Implications for Theory:} By carefully assessing and documenting the performance of LLMs against human programmers across multiple coding characteristics, this research has filled a significant gap in the current literature~\cite{b18,b11,b14,b5}. While previous works have explored LLMs effectiveness at solving programming tasks~\cite{b26,b27} and how they are implemented in developers' workflows~\cite{b30,b31}, limited research has assessed LLMs against humans. This study advanced our awareness of LLMs' ability in supporting software development, offering a fair assessment that takes into account both the drawbacks and possibilities of existing LLM models. However, there is need for additional research into LLMs' capabilities (further considered in Section VIII).\

\textbf{Implications for Practice:} Results in this study have numerous implications for practice. First, we provide evidence that may guide the software engineering community on how LLMs may be integrated into software development processes. Although there is no denying the appeal of automating repetitive coding jobs, this study emphasises the need for strict oversight and review procedures to guarantee that the productivity improvements from utilising LLMs do not compromise the security and quality of the software code generated. Second, our results advance the current discussion concerning applying LLMs to support coding among software engineers. We offer important insights that may improve the deployment methods of LLMs in software development, ensuring that they complement rather than replace human expertise by identifying the precise domains in which they perform well or poorly. These insights will inform software engineers.\

\textbf{Implications for LLM Development and Proofing:} This research not only emphasises how much LLMs may change software development processes, but also how urgently improved training models that put software security and quality first are needed. We need to approach the potential transformation in software development that LLMs may bring about with cautious optimism, recognising the advantages of automation while actively reducing faulty code risks. One approach to enhance LLMs is to equip the LLM training pipeline with specialised training in safe coding techniques. Also, developers may investigate hybrid models, in which LLMs' outputs are placed through a human-in-the-loop review process to improve the generated code and make sure it satisfies the highest standards of quality and security. This way LLMs will provide the most value to software developers.

\section{Threats}
\label{sec:limitations}
 
While our study provides evidence into LLMs and how they perform against human coders, where such evidence is scarce, we concede that the work suffers from a few limitations. Our dataset comprised tasks in relation to one programming language (i.e., Python) and code developed by a fourth year computing student, which is limited in scope and expertise. Thus, our findings may not generalise to other programming languages, where differing syntax and semantic concerns could have a major impact on the security and quality of both human and LLM code generation. In addition, our results may not generalise to situations/domains where the task requirements are significantly different to those of the study. 

In terms of the LLMs, we used GPT-4, which was held to be the best performing LLM at the time of our experiment in April 2024. This is only one LLM, which was no doubt trained on a specific dataset. This may affect the effectiveness of the generated code because LLMs are typically trained on large, diversified datasets that may not particularly focus on secured or optimised coding strategies. This constraint might have biased our comparison and undervalued LLMs' actual capabilities when trained specifically to produce safe, high-quality code. Accordingly, more realistic and balanced comparisons may be obtained by using other LLMs that have been trained with a focus on secure and quality-driven coding techniques, or those that superseded GPT-4 (e.g., GPT-4o or GPT-o1). 
Also, for prompting we used a brief overviews of the task's requirements, expected outcomes, and specific coding guidelines for the LLM to follow (refer to Section III.A). However, an approach that iteratively evolves prompts with errors detected may help to reduce code errors, albeit that would have biased out analysis. 

Finally, we conducted our research using the well-known Pylint, Radon, and Bandit tools. These tools identify a range of problems, however, there may be others that we have missed. For instance, Flake8 combines the functionalities of Ned Batchelder's McCabe, PyFlakes and pycodestyle to provide a large range of metrics. However, this tool does not provide a thorough security analysis, and its capabilities overlap with Pylint and Radon. Also, we have used the guidance of previous work in our consideration of appropriate and relevant checks~\cite{b25}. 

\section{Conclusion}
\label{sec:conclusion}
This research compares the outputs produced by LLMs and human programmers in order to explore the quality of code generated in these contexts and the complexities of software code generation. This study offers detailed understandings into the capabilities and constraints of human and LLMs in the field of software development by investigating their outputs across multiple dimensions, including code compliance to coding standards, security, complexity, and correctness.\

Outcomes revealed that the inclination of LLMs to inherit and duplicate risky coding patterns found in their training datasets is noteworthy. Vulnerabilities including cross-site scripting were significantly more common in code generated by the LLM (i.e., GPT-4) studied. This highlights a crucial flaw in the way that current LLMs' training approaches emphasise learning depth over breadth, highlighting the importance of focused instructions that promote safe coding techniques.

In the coding standards assessment, Pylint and Radon analyses showed that although LLM-generated and human-written code both have flaws, LLMs are more likely to make mistakes that may be avoided by strictly adhering to coding standards. Although modest, the LLM studied (and humans) at times overlooked important details in coding style and standard procedures that are essential for preserving readability and code quality.\

The complexity of the code that was produced was one of the more striking contrasts that was seen. Compared to human-generated code, LLMs often generated code with a higher cyclomatic complexity. Our outcomes suggest that LLMs tend to over-engineer solutions, which could result in code that is harder to maintain and more prone to errors during the later stages of software development.

That said, LLM-generated code frequently performed well when tested for functional correctness, proving LLMs' usefulness in automating simple and well-defined coding tasks. On the contrary, humans frequently outperformed LLMs in tasks requiring deep domain-specific expertise or complicated problem-solving. These outcomes support the view that LLMs should complement (as opposed to replacing) human programmers, where vigilance would go a long way in ensuring the creation of safe and reliable software development outcomes.

\section{Future Work}
In terms of avenues for future studies, we plan to explore LLMs' suitability for solving coding problems in other domains (e.g., for generating solutions for enterprise applications and embedded devices). Also, such analysis may be performed for other programming environments, which would add to the body of knowledge on LLMs' adaptability and versatility. More research is also necessary to determine the moral and security implications of using LLM-generated code in software production, given the vital role that software plays in today's society. Also, creating best practices and frameworks for secure and responsible application of LLMs in software development should be the main focus of future research~\cite{b13}, and we plan to contribute to this research direction. Further, research studies that monitor the growth of LLMs' capabilities over time, especially as new models and training approaches appear, may provide important details concerning the future of LLM technology and its likely important impact on software development~\cite{b12, b15}. We plan such investigations by comparing the performance of other LLMs (e.g., LLaMA and Claude)  when solving the same coding tasks used in this study.\

\label{sec:references}

\end{document}